\documentclass[letterpaper, 10 pt, conference]{IEEEtran}

\IEEEoverridecommandlockouts                              

\usepackage{graphicx} 
\usepackage{amsmath} 

\usepackage{tikz}
\usepackage{array}
\usepackage{booktabs}
\usepackage{subcaption}
\usepackage{textcomp}
\usepackage{rotating}
\usepackage{xcolor}
\usepackage{colortbl}
\usepackage{threeparttable}

\usepackage{diagbox}

\usepackage{algorithm}
\usepackage{algorithmicx}
\usepackage{algpseudocode}
\usepackage{subcaption}
\usepackage{hyperref}

\begin{document}

\title{Bit-Flipping Attack Exploration and Countermeasure in 5G Network\\
}

\author{\IEEEauthorblockN{Joon Kim}
\IEEEauthorblockA{\textit{Department of EECS} \\
\textit{UC Berkeley}\\
Berkeley, CA \\
joonkim1@berkeley.edu}
\and
\IEEEauthorblockN{Chengwei Duan}
\IEEEauthorblockA{\textit{Department of ECE} \\
\textit{University of Florida}\\
Gainesville, FL \\
duan.c@ufl.edu}
\and
\IEEEauthorblockN{Sandip Ray}
\IEEEauthorblockA{\textit{Department of ECE} \\
\textit{University of Florida}\\
Gainesville, FL \\
sandip@ece.ufl.edu}
}


\maketitle
\thispagestyle{empty}
\pagestyle{empty}

\begin{abstract}

5G communication technology has become a vital component in a wide range of applications due to its unique advantages such as high data rate and low latency. While much of the existing research has focused on optimizing its efficiency and performance, security considerations have not received comparable attention, potentially leaving critical vulnerabilities unexplored. In this work, we investigate the vulnerability of 5G systems to bit-flipping attacks, which is an integrity attack where an adversary intercepts 5G network traffic and modifies specific fields of an encrypted message without decryption, thus mutating the message while remaining valid to the receiver. Notably, these attacks do not require the attacker to know the plaintext, and only the semantic meaning or position of certain fields would be enough to effect targeted modifications. We conduct our analysis on OpenAirInterface (OAI), an open-source 5G platform that follows the 3GPP Technical Specifications, to rigorously test the real-world feasibility and impact of bit-flipping attacks under current 5G encryption mechanisms. Finally, we propose a keystream-based shuffling defense mechanism to mitigate the effect of such attacks by raising the difficulty of manipulating specific encrypted fields, while introducing no additional communication overhead compared to the NAS Integrity Algorithm (NIA) in 5G. Our findings reveal that enhancements to 5G security are needed to better protect against attacks that alter data during transmission at the network level.

\end{abstract}

\section{Introduction}

The deployment of the fifth-generation (5G) wireless network marks a transformative leap in communication technology, offering high data rates, ultra-reliable low-latency communications, and massive connectivity. These advancements make 5G a foundational technology for a wide range of emerging applications for Internet of Things (IoT), smart city, industrial automation, etc. \cite{series2015imt} \cite{andrews2014will} Notably, 5G is expected to play a crucial role in enabling connected and autonomous vehicle (CAV) applications \cite{campolo20185g}, where its capabilities can support real-time vehicular communication and safe automated driving operations.

Research on 5G has emphasized improving communication efficiency. However, studies on its security aspects remain comparatively limited. At the network level of 5G (defined as layers above the physical layer but below the application layer) where the data is processed as a digital bitstream, two main security mechanisms are used after authentication: NAS Encryption Algorithm (NEA) for confidentiality and NAS Integrity Algorithm (NIA) for integrity. Although NEA employs strong encryption algorithms and generates a fresh keystream for every packet, it operates via a simple XOR with the plaintext therefore introducing vulnerabilities. Theoretically, a bit-flipping attack, where an adversary can alter specific bits in the ciphertext, can exploit this vulnerability to produce predictable mutations to the plaintext without decryption. NIA integrity checks are effective at detecting such tampering, but it is optional for user-plane data as specified in 3GPP Technical Specifications \cite{TS33501}. This is because NIA introduces additional overhead by appending a 32-bit message authentication code to each packet, increasing both bandwidth usage and processing demands on the transmission entities. Moreover, not all 5G applications require perfect message fidelity. In some cases, occasional isolated mutations have negligible impact. For instance, research shows that in Cooperative Adaptive Cruise Control (CACC) applications, which can be operated under 5G, isolated and discrete mutation attacks on the transmitted acceleration values produce little negative effect compared to the benign, unattacked case \cite{boddupalli2022resilient}. This suggests that for certain 5G applications, mandatory integrity protection with NIA may not be necessary to address bit-flipping attacks.

In this work, we empirically test whether bit-flipping attacks under encryption can truly mutate transmitted values as expected. To achieve this, we avoid building a simplified or toy implementation of 5G network. Instead, we conduct our experiments using OpenAirInterface (OAI), which is a widely used open-source software platform that implements the 5G new radio cellular network closely following the 3GPP Technical Specifications. Additionally, we make use of the one-time pad nature of NEA keystreams to design a keystream-based shuffling algorithm, which significantly reduces the likelihood that a bit-flipping attack could successfully mutate a message while keeping it valid with no additional communication overhead.


\section{Background and Related Work}\label{background}


\subsection{Fundamental of User Data Transmission in 5G}

\begin{figure}[t]
    \centering
    \includegraphics[width=0.9\columnwidth]{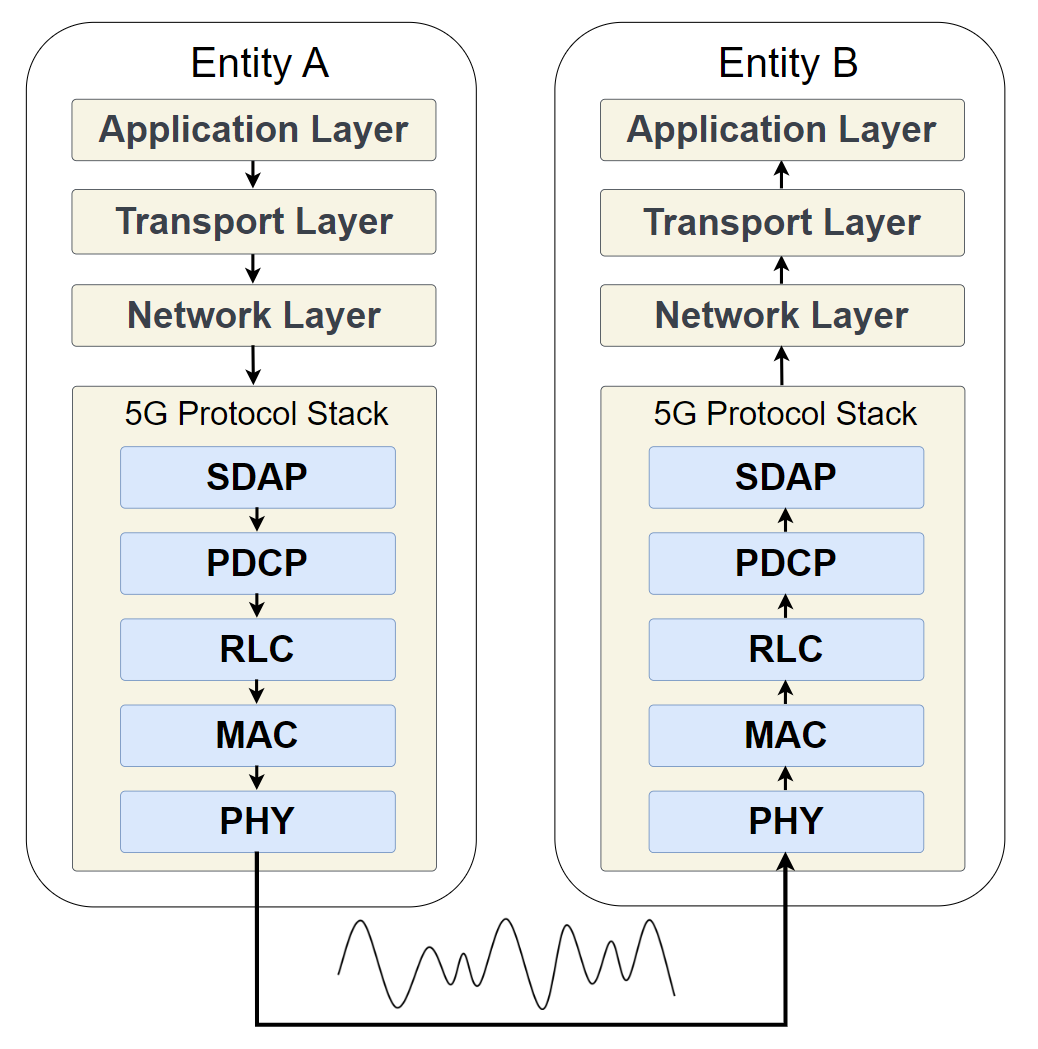}    
    \caption{Layered Architecture of User-plane Data Transmission in 5G}
    \label{fig:DataTrans}
\end{figure}

Currently, most 5G applications still use IP-based communication for compatibility with existing Internet infrastructure and widely used protocols. The overall process is illustrated in Fig. \ref{fig:DataTrans}. When an application needs to transmit some data to others, such as a numeric value, it first converts the data into a bitstream using a standardized format (e.g., IEEE 754 for floating-point numbers). This bitstream then moves down the protocol stack from the transport layer to the 5G protocol stack. At the physical layer, the data is converted into electromagnetic signals and sent over the air to the receiver. On reception, this process is reversed through each layer, ultimately reconstructing the original bitstream so the application layer can recover the transmitted value\footnote{For clarity, this discussion abstracts away certain elements, such as the 5G core network, since they are not central to our research focus.}. 

At the network level, three main mechanisms are employed for safeguarding data confidentiality and integrity during transmission:

\subsubsection{Checksum Error Detection} The transport and network layers use checksum in their headers that allow the receiver to detect errors introduced during transmission. Only the transport layer checksum covers the user data payload to verify the entire segment and protect it. For the transport layer protocols TCP and UDP, the data is divided into 2-byte words for checksum calculation as specified by IETF standards. If the length of the data is not a multiple of 2 bytes, zeros would be padded to the end to complete the final 2-byte word. By summing these 2-byte words and storing the one's complement of the sum, the protocol enables one-sided error detection. Any mismatch at the receiver side indicates that bits were corrupted in transit, while a match does not necessarily guarantee the absence of corruption.

\subsubsection{Encryption Algorithm for 5G (NEA)} 
\begin{figure}[t]
    \centering
    \includegraphics[width=1.0\columnwidth]{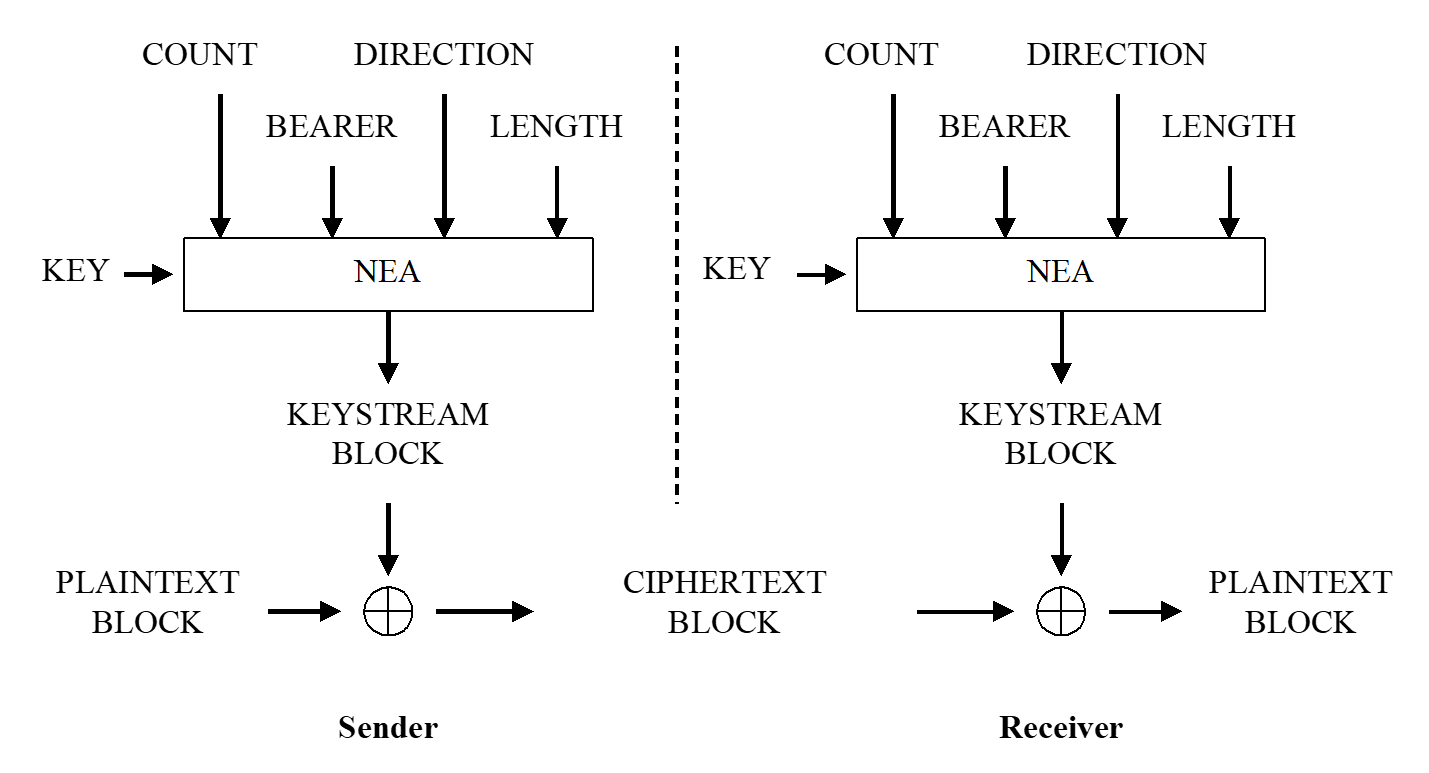}    
    \caption{5G Ciphering Process}
    \label{fig:NEA}
\end{figure}

Confidentiality of data transmitted over 5G is enforced at the PDCP layer using the NEA algorithm, as shown in Fig. \ref{fig:NEA} \footnote{Fig ~\ref{fig:NEA} is adapted from 3GPP TS 33.501 \cite{TS33501}.}. On the sender side, NEA uses the secret key established during the 5G authentication and key agreement process (AKA), and control parameters derived from synchronized protocol state or unencrypted framing to generate a keystream. This keystream is then XORed with the plaintext bitstream to produce the ciphertext. On the receiver's side, the same keystream is generated using the same key and synchronized parameters, and XORed again with the ciphertext to recover the original plaintext. This mechanism ensures that a third party cannot understand the intercepted data without access to the secret key and relevant control information.

\subsubsection{Integrity Algorithm for 5G (NIA)} Integrity protection in 5G is achieved using the NIA algorithm. The sender generates a 32-bit message authentication code (MAC-I) by applying NIA to the message, key, and relevant control parameters, and appends this MAC to the transmitted data. Upon reception, the receiver uses the same key and parameters to compute the expected MAC (after decryption if encryption is used) and compares it to the received MAC-I. A match confirms data integrity, while a mismatch indicates possible tampering or transmission errors, prompting the receiver to discard the message. This method prevents unauthorized modifications to the data at the cost of adding a 4-byte authentication tag to each transmission.

\subsection{Related Work}


Bit-flipping attacks have been studied in various network environments. Paterson et al. demonstrated bit-flipping attacks on IPSec-protected datagrams and analyzed their impacts \cite{paterson2006cryptography}. The same type of attacks has also been examined in Long-Range Wide-Area Network (LoRaWAN). Lee et al. conducted a risk analysis of bit-flipping in LoRaWAN and proposed a key-based shuffling mechanism using circular shifts and swaps to mitigate such threats \cite{lee2017risk}. In the context of cellular networks, Rupprecht et al. introduced the ALTER attack on LTE, which manipulates data payloads by adding a manipulation mask that selectively flips bits in the message. In their work, altering fields such as the destination IP address required the attacker to know the original plaintext to compute the correct manipulation mask, leading the authors to argue that robust, mandatory integrity protection in 5G is a necessary countermeasure \cite{rupprecht2019breaking}. Tan et al. discussed bit-flipping attacks on 5G data plane packets, focusing on parts of the message with predictable content, such as IP headers or retransmission indicators. They similarly noted that an attacker typically needs to know the original plaintext to forge a targeted modification via bit-flipping \cite{tan2022system}. However, our study challenges this assumption, showing that knowledge of the message format structure alone is often sufficient to conduct successful bit-flipping attacks and achieve predictable results. Additionally, previous bit-flipping research in 4G and 5G has primarily targeted control fields where the plaintext is known, rather than exploring the potential impact of bit-flipping attacks on less predictable data payloads, which is an area that our work addresses.

\section{Bit-Flipping Attack in 5G} \label{attack}

\subsection{Threat Model}

We define key terms for bit-flipping attacks and the adversary capabilities in this subsection. In a 5G communication setting, let $\mathcal{S}$ denote the sender, $\mathcal{R}$ denote the receiver, and $\mathcal{A}$ denote the adversary. In a benign scenario, $\mathcal{S}$ transmits a message over 5G, and $\mathcal{R}$ reliably receives the correct message. However, the addition of $\mathcal{A}$ no longer guarantees reliable communication as $\mathcal{A}$'s goal is to mutate as many messages as possible. We assume the followings for our threat model:

\begin{itemize}
    \item $\mathcal{A}$ acts as a Man-in-the-Middle (MITM) attacker that is able to intercept the physical-layer signal from $\mathcal{S}$ to $\mathcal{R}$.
    \item $\mathcal{A}$ possesses the capability to decode the intercepted signal, reconstruct the PDCP-layer ciphertext bitstream, and flip any number of bits at arbitrary positions correspond to the field of checksum and data payload.
    \item After mutation, $\mathcal{A}$ can re-encode and forward the modified message to $\mathcal{R}$.
    \item $\mathcal{A}$ cannot decrypt the NEA encrypted ciphertext.
\end{itemize}

\subsection{Bit-Flipping Attack}



\begin{figure}[t]
    \centering
    \includegraphics[width=1.0\columnwidth]{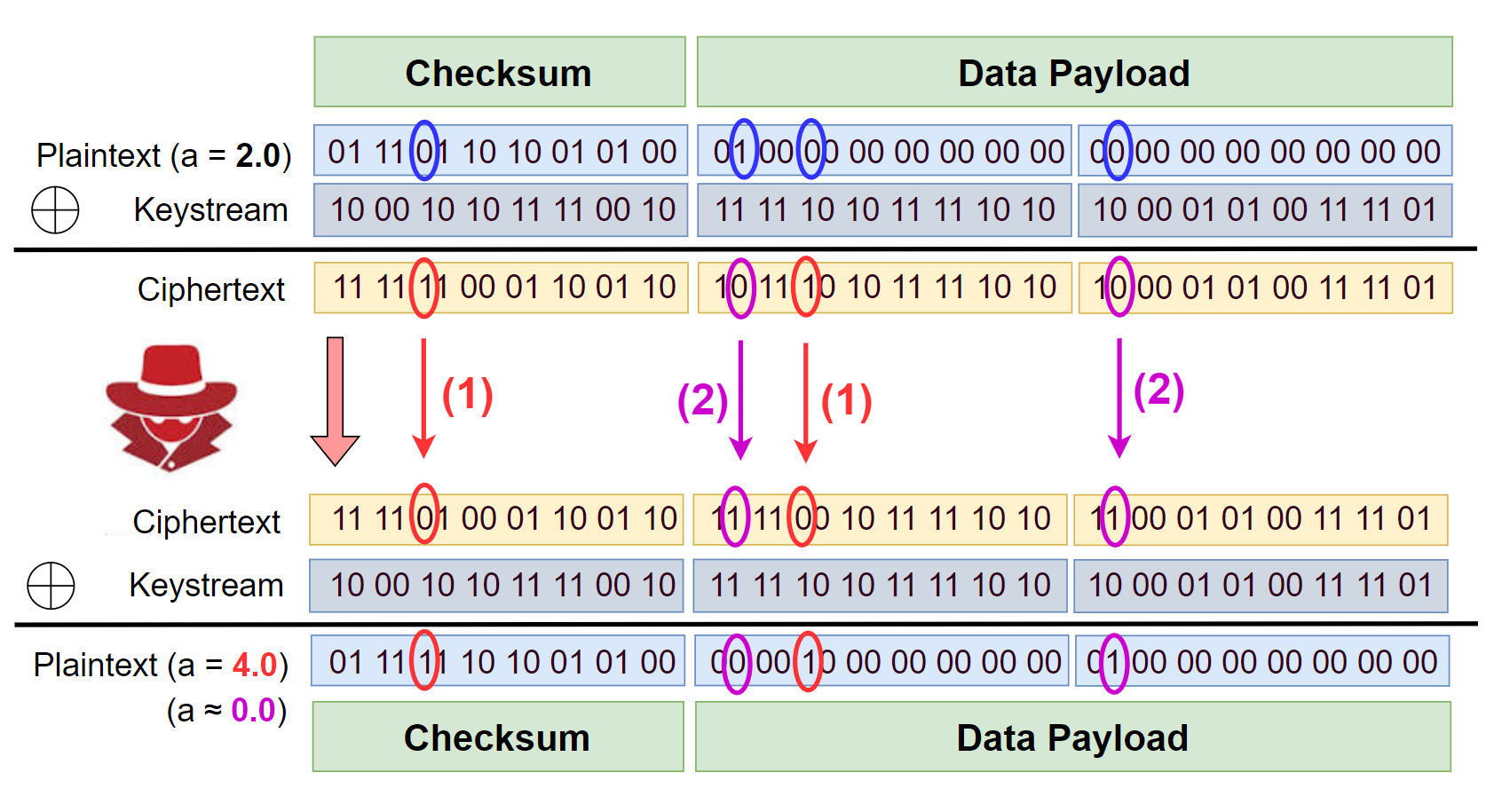}    
    \caption{Bit-Flipping Attack}
    \label{fig:bitflipattack}
\end{figure}

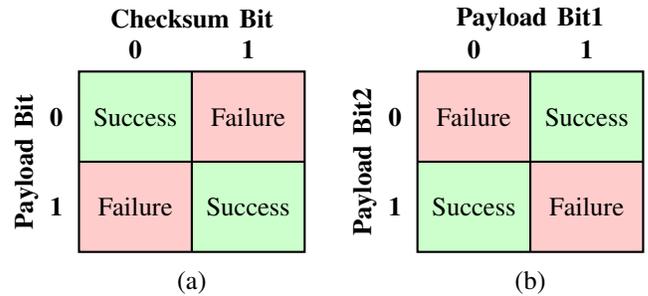
\begin{figure}[t]
\centering
\begin{tikzpicture}

    
    \fill[green!20] (-3.5,1.2) rectangle (-2,2.4);
    \fill[red!20] (-2,1.2) rectangle (-0.5,2.4);
    \fill[red!20] (-3.5,0) rectangle (-2,1.2);
    \fill[green!20] (-2,0) rectangle (-0.5,1.2);
    
    \draw[thick] (-3.5,0) rectangle (-0.5,2.4);
    \draw[thick] (-3.5,1.2) -- (-0.5,1.2);
    \draw[thick] (-2,0) -- (-2,2.4);
    
    \node at (-2.75,1.8) {Success};
    \node at (-1.25,1.8) {Failure};
    \node at (-2.75,0.6) {Failure};
    \node at (-1.25,0.6) {Success};
    
    \node at (-2.75,2.7) {\textbf{0}};
    \node at (-1.25,2.7) {\textbf{1}};
    
    \node at (-3.8,1.8) {\textbf{0}};
    \node at (-3.8,0.6) {\textbf{1}};
    
    \node at (-2,3.1) {\textbf{Checksum Bit}};
    
    \node[rotate=90] at (-4.2,1.2) {\textbf{Payload Bit}};

    \node at (-2,-0.4) {(a)};


    \fill[red!20] (1,1.2) rectangle (2.5,2.4);
    \fill[green!20] (2.5,1.2) rectangle (4,2.4);
    \fill[green!20] (1,0) rectangle (2.5,1.2);
    \fill[red!20] (2.5,0) rectangle (4,1.2);

    \draw[thick] (1,0) rectangle (4,2.4);
    \draw[thick] (1,1.2) -- (4,1.2);
    \draw[thick] (2.5,0) -- (2.5,2.4);

    \node at (1.75,1.8) {Failure};
    \node at (3.25,1.8) {Success};
    \node at (1.75,0.6) {Success};
    \node at (3.25,0.6) {Failure};

    \node at (1.75,2.7) {\textbf{0}};
    \node at (3.25,2.7) {\textbf{1}};

    \node at (0.7,1.8) {\textbf{0}};
    \node at (0.7,0.6) {\textbf{1}};

    \node at (2.5,3.1) {\textbf{Payload Bit1}};

    \node[rotate=90] at (0.3,1.2) {\textbf{Payload Bit2}};

    \node at (2.5,-0.4) {(b)};
    
\end{tikzpicture}
\caption{Success-Failure Matrix of (a) Checksum Bit-Flipping Attack, (b) Payload Bit-Flipping Attack. The two flipped bits must be aligned in the same column when divided into two-byte chunks for checksum calculation.}
\label{fig:2bitflip}
\end{figure}

Even if $\mathcal{A}$ cannot decrypt the message or learn its plaintext, it can still strategically flip selected bits to maximize its chance to bypass the checksum error detection. Consider the bit-flipping attack illustrated by the red arrows (1) in Fig. \ref{fig:bitflipattack}: $\mathcal{S}$ sends a value 2.0 to $\mathcal{R}$. $\mathcal{A}$ flips two bits, one in the checksum field, and the other in the data payload field that aligns with it in the 2-byte words. The attack will successfully bypass the checksum verification and mutate one bit in the data payload if and only if the parity of the two flipped bits is even in plaintext, as illustrated in Fig. \ref{fig:2bitflip} (a). The plaintext is then deciphered to a value of 4.0. This yields an attack success rate around 50\% for any message since the checksum field is almost pairwise independent of any one aligned bit in the data payload. Such independence arises because checksum computation incorporates not just the data payload, but also other fields like source and destination ports, message length, etc., which may vary across applications. We refer to such attacks as “\textit{Checksum Bit-Flipping Attacks}”. Another attack strategy is to flip two aligned bits within the data payload, which we call “\textit{Payload Bit-Flipping Attacks}” illustrated by the pink arrows (2) in Fig. \ref{fig:bitflipattack}. In this case, the attack succeeds when the parity of the flipped bits is odd in plaintext, as illustrated in Fig. \ref{fig:2bitflip} (b), and the value is mutated to a small value close to 0.0. While this allows $\mathcal{A}$ to mutate two bits, its success depends directly on the underlying plaintext values, and the attack does not benefit from the pairwise independence property seen in the checksum bit-flipping case.

As a MITM attacker, there is no limitation on the number of bits $\mathcal{A}$ could flip. The only issue is that there is a stark trade-off between the number of bit flips and the mutation success rate. For checksum bit-flipping attacks on separate checksum bits, we empirically show that there is an exponentially decaying trend of success rates with respect to the number of flipped bits. Payload bit-flipping attacks are difficult to analyze due to their high dependence on the specific value of the data payload plaintext. Nonetheless, we expect them to also deteriorate in success rate quickly as the number of flips increase.



\section{Keystream-Based Shuffling Defense} \label{defense}


In this section, we propose a general defense mechanism against bit-flipping attacks via shuffling. We first observe that the plaintext and the ciphertext have the same columnwise alignment; i.e. flipping a position in the ciphertext directly corresponds to flipping that same position in the plaintext itself. However, if the ciphertext is shuffled unpredictably, $\mathcal{A}$ would have much less confidence in being able to pinpoint specific positions to attack. The challenge is to ensure that $\mathcal{A}$ cannot deduce the mechanism of the shuffling while enabling $\mathcal{R}$ to perform the inverse of the shuffle deterministically.


Theoretically, this is impossible to achieve without establishing some form of private information between $\mathcal{S}$ and $\mathcal{R}$. Fortunately, 5G NEA lets $\mathcal{S}$ and $\mathcal{R}$ generate a unique, secret keystream for each message transmitted between them. The idea is to then reuse the private keystream generated for a message to seed a pseudorandom permutation function that shuffles the checksum and data payload fields. $\mathcal{A}$, without the knowledge of the cipher key and keystream, would be cryptographically unable to guess each bit's original position better than random. On the contrary, it is easy for $\mathcal{R}$ to construct an inverse shuffling procedure to obtain the original plaintext.

\begin{algorithm}[t]
\caption{Keystream-based Shuffling Implementation}
\label{alg:sender_receiver}
\begin{algorithmic}[1]
\Statex \textbf{Shared:} Cipher Key $k$, control parameters $paras$ for each transmit message, Pseudorandom Permutation Generator $PRP()$

\Statex

\Procedure{Sender}{Plaintext $P$}
    \State $K \gets NEA(k, paras)$  \Comment{Generate Keystream} 
    \State $C \gets P \oplus K$ \Comment{XOR Ciphering} 
    \State $T \gets PRP(K)$ \Comment{Permutation Table}
    \State create $C_s$ s.t. $|C_s| = |C|$
    \For{$i$ in $len(C)$}
        \State $C_s[T[i]] = C[i]$ \Comment{Shuffle according to $T$}
    \EndFor
    \State $send(C_S)$
\EndProcedure

\Statex  

\Procedure{Receiver}{Shuffled Ciphertext $C_s$}
    \State $K \gets NEA(k, paras)$ 
    \State $T \gets PRP(K)$
    \State create $C_u$ s.t. $|C_u| = |C_s|$
    \For{$i$ in $len(C_s)$}
        \State $C_u[i] = C_s[T[i]]$ \Comment{Invert the shuffle}
    \EndFor   
    \State $P \gets C_u \oplus K$ \Comment{XOR$(\cdot, K)$ is its own inverse}
\EndProcedure

\end{algorithmic}
\end{algorithm}



Our defense mechanism leverages the Fisher-Yates shuffling algorithm, using the keystream generated for each transmission to determine the shuffle order, and applies this process directly to the ciphertext. The receiver would construct the forward permutation table with the same keystream, then invert it to produce the inverse permutation table. The plaintext is extracted after reversing the effects of the shuffle. Algorithm \ref{alg:sender_receiver} generalizes the implementation over different choices of pseudorandom permutation generators.

The keystream-based shuffling proposes an alternative defense mechanism to NIA, and there are inherent trade-offs. NIA is a strong integrity protection algorithm that can detect any tampering of the data payload with near zero failure. However, it requires redundant bytes for each message. Shuffling, on the other hand, does not append any additional bits to the transmitted message, gaining an advantage in latency and communication overhead. The sacrifice is that shuffling is not deterministic and has a chance to fail to defend. Theoretically, for 2-bit flips, it is expected that the attack still succeeds with probability close to $\frac{1}{16} \cdot \frac{1}{2} = \frac{1}{32} = 0.03125$. This is because there is only a $\frac{1}{16}$ chance that two randomly selected positions will be aligned in the two bytes for checksum, and if we reasonably assume that no two bits are highly correlated with each other, there is a uniform $\frac{1}{2}$ chance that they match in parity. While not perfect, it is still a significant improvement over deactivating NIA and only using checksum, which has a $\frac{1}{2}$ failure rate. For applications such as CACC, we expect that such low mutation success rate will be enough to disallow contiguous attacks for extended numbers of timesteps. Finally, shuffling has the unique trait that it also obfuscates the position of the bits, effectively randomizing the position of the bit flips attempted by $\mathcal{A}.$ Even if $\mathcal{A}$ successfully mutates a payload, it has no control over which bit was actually attacked, which eliminates its capabilities to generate intentional attacks on specific bit fields.

\section{Experiment Results} \label{results}


In this section, we first assess the feasibility of bit-flipping attacks on 5G networks using the OAI platform. Then, to evaluate practical implications, we simulate a scenario in which vehicle A communicates its position, velocity, and acceleration to vehicle B through 5G. For this purpose, we select three vehicles at random from the Next Generation Simulation (NGSIM) dataset, collected by the U.S. Department of Transportation \cite{ngsim2016}, and use their real-world trajectories as the shared states. We then analyze the impact of checksum bit-flipping attacks and the mitigation performance of our proposed shuffling-based defense in this context. Finally, we extend our analysis to examine the effects of payload bit-flipping attacks on a fourth vehicle.

\subsection{Bit-Flipping Attacks Simulation}

We simulate the bit-flipping attacks using OAI, a reputable open-source simulation software platform used for 5G networking research. Since our threat model assumes that the attacker $\mathcal{A}$ can intercept and decode transmissions to the ciphertext level, we implement the bit-flipping attacks on the sender side for convenience. The attacks are implemented by modifying the OAI source code in the \textit{openair2/LAYER2/nr\_pdcp} directory, specifically by editing the \textit{deliver\_pdu\_drb\_ue} function in \textit{nr\_pdcp\_oai\_api.c}. Our keystream-based shuffling defense is also implemented at the PDCP layer in OAI, with modifications to the functions \textit{nr\_pdcp\_entity\_process\_sdu} and \textit{nr\_pdcp\_entity\_recv\_pdu} in \textit{nr\_pdcp\_entity.c}\footnote{The code associated with this paper can be found at \url{https://github.com/DerekDuan615/5G-Bit-Flipping-Attack-Exploration} (branch: keystream-based-shuffle).}.

\begin{figure*}
    \centering
    \includegraphics[width=0.8\linewidth]{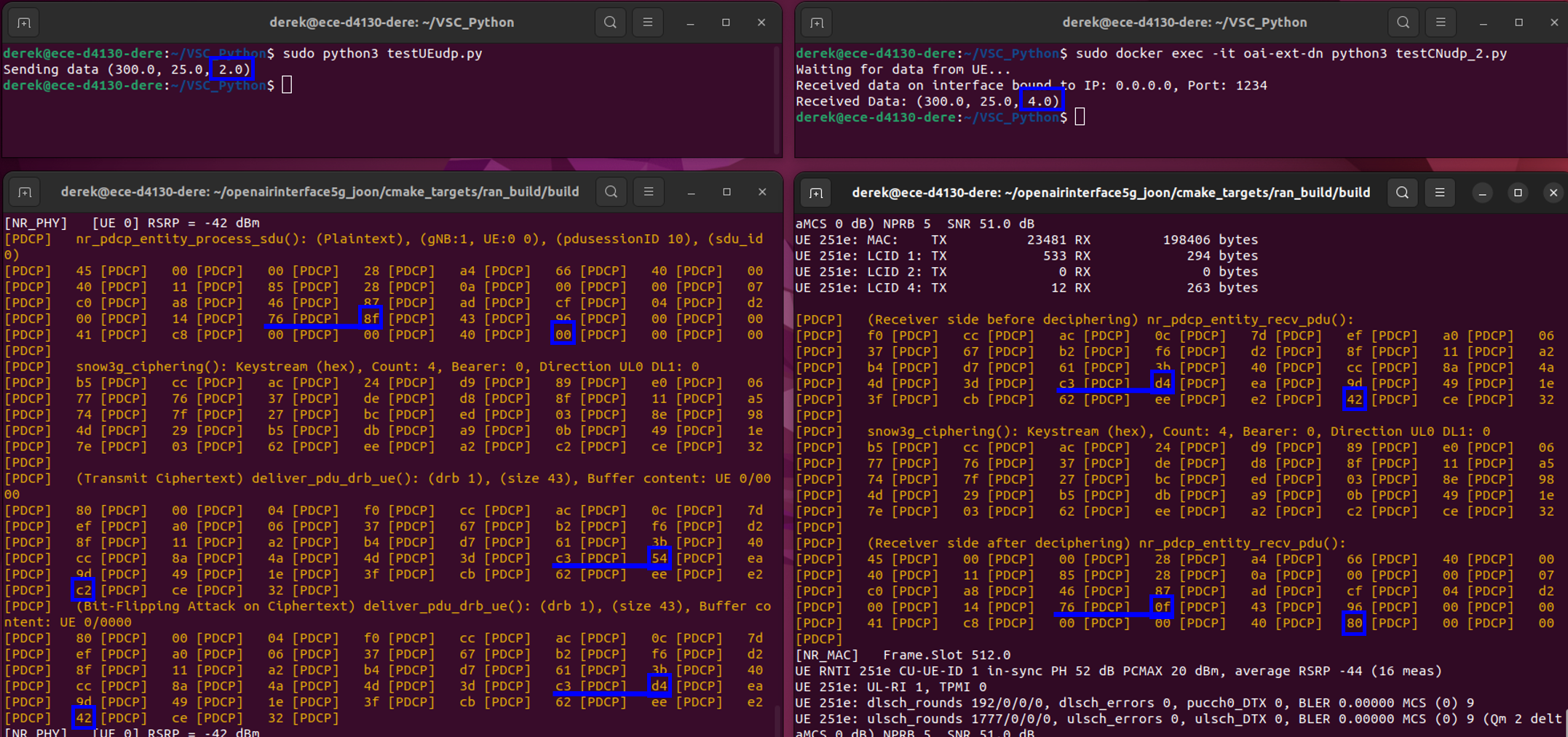}
    \caption{Checksum Bit-Flipping Attack (left: vehicle A; right: vehicle B)}
    \label{fig:CheckBitFlip}
\end{figure*}

\begin{figure*}
    \centering
    \includegraphics[width=0.8\linewidth]{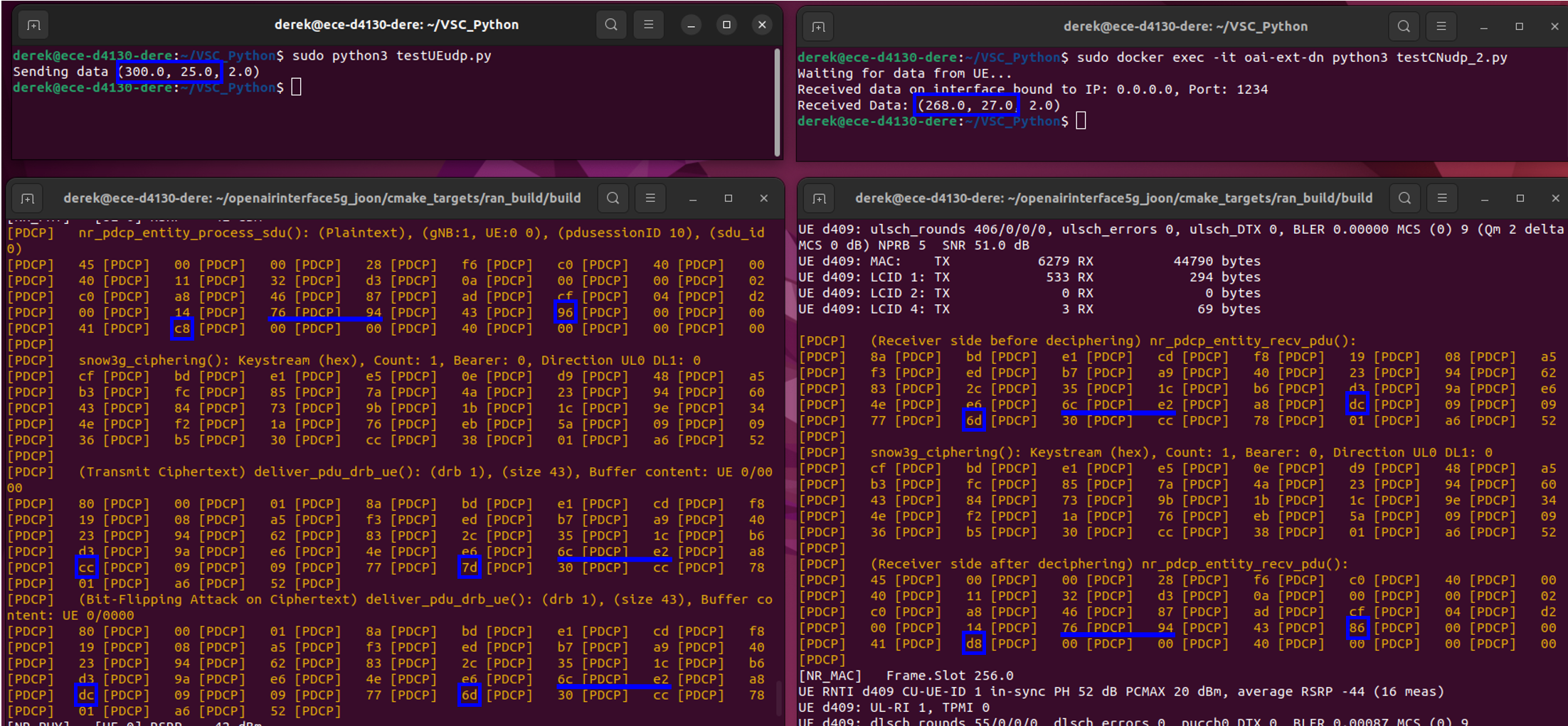}
    \caption{Payload Bit-Flipping Attack (left: vehicle A; right: vehicle B)}
    \label{fig:PayBitFlip}
\end{figure*}

We verify the effectiveness of bit-flipping attack in a scenario where vehicle A sends its position, velocity, and acceleration (set to 300.0, 25.0, and 2.0) to vehicle B. Results for both checksum and payload bit-flipping attacks are shown in Fig. \ref{fig:CheckBitFlip} and Fig. \ref{fig:PayBitFlip}, with blue dashes marking checksum locations and blue boxes indicating bit-flip and mutated values. In the checksum bit-flipping attack, the adversary alters the ciphered checksum field from \texttt{0xc354} to \texttt{0xc3d4}, and the ciphered acceleration value from \texttt{0xe2c2ce32} to \texttt{0xe242ce32}. As a result, the receiver decrypts the mutated message, which passes the checksum verification and results in a received acceleration value of 4.0, as displayed in the upper-right terminal in Fig. \ref{fig:CheckBitFlip}. It is important to note that OAI’s checksum implementation inverts all checksum bits after calculation, so the observed success-failure pattern is exactly the opposite of what is shown in Fig. \ref{fig:2bitflip} (a). For the payload bit-flipping attack, the ciphertexts corresponding to the position and velocity values are manipulated, causing the receiver to obtain position and velocity values of 268.0 and 27.0, respectively, which align with the success-failure matrix expected in Fig. \ref{fig:2bitflip} (b).

\begin{figure}[t]
    \centering
    \includegraphics[width=0.8\linewidth]{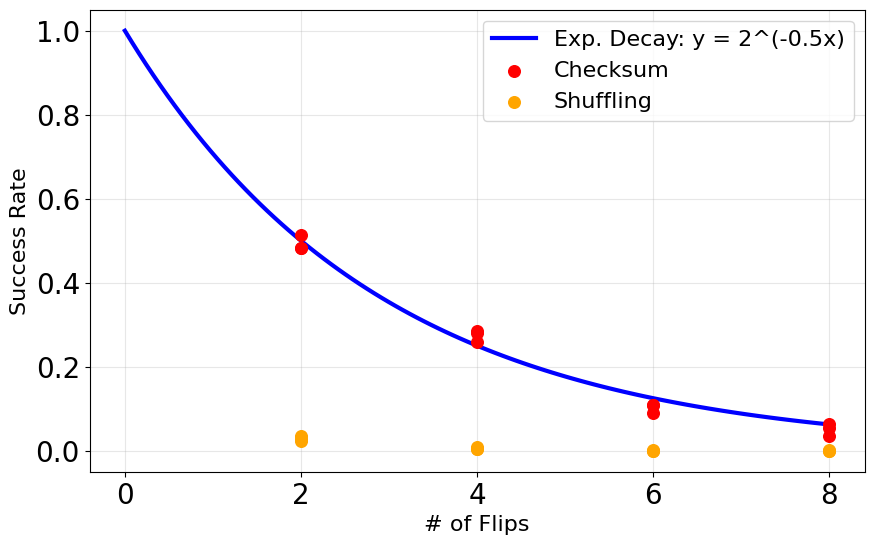}
    \caption{Mutation Success Rate of Checksum Bit-Flipping Attack with and without Shuffling Defense}
    \label{fig:mutation-graph}
\end{figure}

\subsection{Attack and Defense Effect Analysis}

To ensure that our approach targets realistic environments, we used the NGSIM Open Data for vehicle trajectories. Each vehicle's acceleration, velocity, and X-coordinate data were extracted and used as the transmitted message for a single timestep. Each individual experiment consisted of the success rate of 500 data points of a single vehicle.

\begin{table}[t]
\centering
\caption{Mutation Success Rate of Checksum Bit-Flipping Attacks}
\begin{tabular}{|c|c|c|c|c|}
\hline
\diagbox{Trajectory}{\# of flips} & 2-bits & 4-bits & 6-bits & 8-bits \\
\hline
Vehicle1 & 0.514 & 0.280 & 0.110& 0.064 \\
\hline
Vehicle2 & 0.482 & 0.258 & 0.090 & 0.034 \\
\hline
Vehicle3 & 0.482 & 0.284 & 0.108 & 0.054 \\
\hline
\end{tabular}
\label{tab:checksum-defense}
\end{table}

\begin{table}[t]
\centering
\caption{Mutation Success Rate of Checksum Bit-Flipping Attacks with Shuffling Defense}
\begin{tabular}{|c|c|c|c|c|}
\hline
\diagbox{Trajectory}{\# of flips} & 2-bits & 4-bits & 6-bits & 8-bits \\
\hline
Vehicle1 & 0.036 & 0.004 & 0.000 & 0.002 \\
\hline
Vehicle2 & 0.022 & 0.008 & 0.000 & 0.000 \\
\hline
Vehicle3 & 0.028 & 0.004 & 0.002 & 0.000 \\
\hline
\end{tabular}
\label{tab:shuffling-defense}
\end{table}

\begin{table}[t]
\centering
\caption{Mutation Success Rate of 2-bit Flipping Attacks at Varying Payload Positions}
\begin{tabular}{|c|c|}
\hline
\diagbox{Locations}{Trajectory} & Vehicle 4 \\
\hline
Checksum and Acc & 0.530 \\
\hline
Acc, Vel Sign & 0.412 \\
\hline
Acc, Vel Exp & 0.276 \\
\hline
Acc, Vel Mant & 0.436 \\
\hline
\end{tabular}
\label{tab:payload}
\end{table}

Our main experimental results for checksum bit-flipping attacks are summarized in Figure \ref{fig:mutation-graph} and numerically reported in Tables \ref{tab:checksum-defense} and \ref{tab:shuffling-defense}. The red dots show the success rate of a checksum bit-flipping attack when NIA is disabled and the checksum is the sole defense against bit-flipping. While we cannot say that the relationship between the payload and checksum bits being flipped are truly pairwise independent, our experiment empirically proves the exponential decaying trend for the mutation success rate in checksum bit-flipping attacks. In practice, we can estimate that every two additional flips  halves the mutation success rate, following the blue line on the graph ($y = 2^{-x/2}$) very closely. Shuffling, on the other hand, maintains a low mutation success rate even when the number of bit flips is only 2. Indeed, the empirical success rate is close to the theoretical probability of 0.03125 we suggested in the previous section. The success rate rapidly declines as the number of flips increase, and for more than 6 flips, the success rate is almost negligible.


As proof for the non-independence of the payload attacks, we run multiple payload bit-flipping attacks that vary in location. Specifically, we attack the sign bit, the second most significant bit (MSB) of the exponent field, and the MSB of the mantissa field in both acceleration and velocity floats. A checksum bit-flipping attack is added for comparison. As expected, only the checksum bit-flipping attack shows mutation success rate near 50\%, whereas other three attacks hover significantly below. Nonetheless, this does not suggest that payload bit-flipping attacks should always have lower success rate than the checksum attack; the values might as well have been far exceeding 50\%. There might be potential to exploit this fact to increase the attack success rate beyond 50\%. However, this discussion is out of scope for this study.

\section{Conclusion} \label{conclusion}


We identify a feasible MITM bit-flipping attack in 5G networks and provide an alternative shuffling defense mechanism that does not require the redundancy bit overhead with the cost of slightly increased probability of failure compared to NIA. Our key findings through experiments verify that bit-flipping acts according to theoretical expectations and can be successful at mutating one bit with 50\% probability if applied in particular positions. However, since the attack depends on knowing the exact alignment of the bits with respect to the checksum, shuffling effectively disallows the adversary to consistently attack particular positions of the data payload. Ultimately, shuffling is an appealing choice for 5G applications such as CACC, where the small defense failure probability is worth to trade for the additional overhead incurred by NIA. Future work would include analyzing the impact of bit-flipping attacks in CACC applications and developing more consistent yet robust defense mechanisms than shuffling.

\subsubsection*{Acknowledgements}

This work was partially supported by NSF Grants REU-2150136 and SATC-2221900.









\bibliographystyle{IEEEtran}
\bibliography{reference}

\end{document}